\documentclass[twocolumn]{aastex63}

\usepackage{amsmath}

\shorttitle{GRB 210121A: the intermediate photosphere and the evolution of the outflow}
\shortauthors{Xin-Ying Song et al.}

\begin{document}

\title{ GRB 210121A: Observation of photospheric emissions from different regimes and the evolution of the outflow }

\correspondingauthor{Xin-Ying Song}
\email{songxy@ihep.ac.cn}

\author{Xin-Ying Song}
\affiliation{Key Laboratory of Particle Astrophysics, Institute of High Energy Physics, Chinese Academy of Sciences, Beijing 100049, China}
\author{Shuang-Nan Zhang}
\affil{Key Laboratory of Particle Astrophysics, Institute of High Energy Physics, Chinese Academy of Sciences, Beijing 100049, China}
\affil{University of Chinese Academy of Sciences, Chinese Academy of Sciences, Beijing 100049, China}
\author{Shu Zhang}
\affil{Key Laboratory of Particle Astrophysics, Institute of High Energy Physics, Chinese Academy of Sciences, Beijing 100049, China}
\author{Shao-Lin Xiong}
\affil{Key Laboratory of Particle Astrophysics, Institute of High Energy Physics, Chinese Academy of Sciences, Beijing 100049, China}
\author{Li-Ming Song}
\affil{Key Laboratory of Particle Astrophysics, Institute of High Energy Physics, Chinese Academy of Sciences, Beijing 100049, China}





\begin{abstract}
GRB 210121A was observed by Insight-HXMT, Gravitational wave high-energy Electromagnetic Counterpart All-sky Monitor (GECAM), Fermi Gamma-ray Burst Monitor (Fermi/GBM) on Jan 21st 2021. In this work, photospheric emission from a structured jet is preferred to interpret the prompt emission phase of GRB 210121A and emissions from different regimes are observed on-axis. Particularly, the emission from the intermediate photosphere is first observed in the first 1.3 s of the prompt emission, while those from the other part are dominant by the saturated regime, and offers an alternative explanation compared with the previous work. Moreover, the emissions with considering the intermediate photosphere can well interpret the changes on low-energy photon index $\alpha$ during the pulses. Besides, the evolution of the outflow is extracted from time-resolved analysis, and a correlation of $\Gamma_0 \propto L^{0.25\pm0.05}_0$ is obtained, which implies that the jet may be mainly launched by neutrino annihilation in a hyper-accretion disk. 
\end{abstract}

\keywords{Gamma-Ray Bursts, Photosphere, prompt emission}


\section{Introduction} \label{sec:intro}
Despite about a half century of observations, prompt gamma-ray burst (GRB) emission mechanisms are still a matter of interest. \textbf{Two leading scenarios have been suggested to interpret the observed spectra of GRBs. One is synchrotron radiation, which invokes a non-thermal emission of relativistic charged particles either from internal shocks or from internal magnetic dissipation processes \citep{2000ApJ...543..722L,2000AIPC..526..185T,2004ApJ...613..460B,2018pgrb.book.....Z}. 
The fast cooling problem in relevance to the low-energy photon index $\alpha$ could be relieved by synchrotron radiation of electrons in a moderately fast-cooling regime \citep{2014NatPh..10..351U} or detailed treatment of the cooling electrons~\citep{2001A&A...372.1071D, 2018ApJS..234....3G}.}
As a natural consequence of the fireball model, photospheric emission produced from highly relativistic outflows was previously considered as an explanation for prompt gamma-ray bursts~\citep[]{1986Are,1986ApJ308L43P}, where the optical depth at the base of the outflow is much larger than unity~\citep{1999PhR...314..575P}. The Planck spectrum related to the photospheric emission could be broadened in two ways. First, dissipation below the photosphere can heat electrons above the equilibrium temperature. These electrons emit synchrotron emission and comptonize the thermal photons, thereby modify the shape of Planck spectrum~\citep{2005ApJ...635..476P,Pe_er_2006,2005Dissipative}. Observational evidence for subphotospheric heating has been provided by \cite{2011Observational}. Besides, internal shocks below the photosphere~\citep{2005Dissipative}, magnetic reconnection~\citep{1994A,2004Spectra}, and hadronic collision shocks~\citep{2010Collisional,2010Radiative} can also cause dissipation. Second, the modification of Planck spectrum could be caused by geometrical broadening. Photospheric radius is found to be a function of the angle to the line of sight of the photons of thermal emission observed~\citep{1991The,2008ApJ...682..463P}. This means that the observed spectrum is a superposition of a series of blackbodies of different temperature, arising from different angles to the line of sight. Moreover, \cite{2008ApJ...682..463P} showed that photons make their last scatterings at a distribution of radii and angles. The observer sees simultaneously photons emitted from a large range of radii and angles. Therefore, the observed spectrum is a superposition of comoving spectra~\citep{2010ApJ...709L.172R,Hou_2018}. \cite{2013A} studied the non-dissipative photospheric (NDP) emissions from a structured jet, and reproduced the average low-energy photon index ($\alpha=-1$) independent of viewing angle. The observed evolution patterns of the $\nu F_{\nu}$ peak energy ($E_{\rm p}$), including hard-to-soft and intensity-tracking could be reproduced by this model as well~\citep[e.g.][]{Deng_2014,2019The}.


There are three regimes to be discussed in photospheric emission from a structured jet:  (I) unsaturated emissions which are dominant by the regime of unsaturated acceleration (the saturation radius is greater than the photonsphere radius, $R_{\rm s}>R_{\rm ph}$); (II) saturated emissions from the regime of $R_{\rm s}\leq R_{\rm ph}$ works for all over the wind profile; (III) the intermediate photosphere~\citep{2022MNRAS.512.5693S} which represents the case where the regimes of $R_{\rm s}>R_{\rm ph}$ and $R_{\rm s}\leq R_{\rm ph}$ work in lower and higher latitudes respectively, and the contribution from the latter can not be ignored. It has been never mentioned in the previous study in the spectral fit to photospheric emissions. Besides of the off-axis NDP model in unsaturated regime~\citep{Wang_2021}, in this paper, we find that the on-axis NDP model with considering intermediate photosphere is an alternative description for the emissions of GRB 210121A.

 The evolution of the outflow is also extracted, to offer an interpretation about the mechanism for launching the jet. If we take the hyper-accreting black hole (BH) as the central engine, the GRB jet may be launched from two mechanisms. One is $\nu\overline{\nu}$ annihilation in a neutrino-dominated accretion flow (NDAF) \citep[e.g.][]{1999ApJ...518..356P,2001ApJ...557..949N,Di_Matteo_2002,2002ApJ...577..311K,2006ApJ...643L..87G,2007ApJ...657..383C,2007ApJ...664.1011J,2009ApJ...700.1970L,2010A&A...516A..16L}, and generates a fireball which is dominant by the thermal component. The jet is launched by neutrino annihilation ($\nu \overline{\nu}\rightarrow e^{+} e^{-}$), and $\dot{E}_{\nu\overline{\nu}}$ is the neutrino annihilation power. \textbf{The other one is Blandford\&Znajek mechanism~\cite[BZ,][]{1977MNRAS.179..433B}}. The spin energy of the BH is tapped by a magnetic field, and produces a Poynting flux. The correlations between the baryon loading parameter and the power for these two mechanisms are both positive, $\eta\propto\dot{E}^{0.26}_{\nu\overline{\nu}}$~\citep{L_2012} for the former, and $\mu_0\propto\dot{E}^{0.17}_{\rm BZ}$~\citep{YI20171} ($\mu_0=\eta(1+\sigma_0)$, where $\sigma_0$ is the ratio of Poynting flux luminosity to the matter flux) for the latter. Thus, the index of the correlation offers a criterion.

This paper is organized as follows. In Section~\ref{sec:obs_GRB210121A}, the observation of GRB 210121A by different missions is introduced. In Section~\ref{sec:NDPmodel}, the NDP model and $R_{\rm ph}$ in different regimes are introduced, especially in the intermediate photosphere. In Section~\ref{sec:TRana}, the methods for binning, background estimation and the spectral fitting are clarified. In Section~\ref{sec:NDPfit}, time-resolved analyses are performed; the properties of the jet are extracted, while the regimes are determined. Then the discussion and conclusion are given in Section~\ref{sec:discuss}.

\section{observation of GRB210121A}\label{sec:obs_GRB210121A}

\begin{figure*}
\begin{center}
 \centering
  \includegraphics[width=\textwidth]{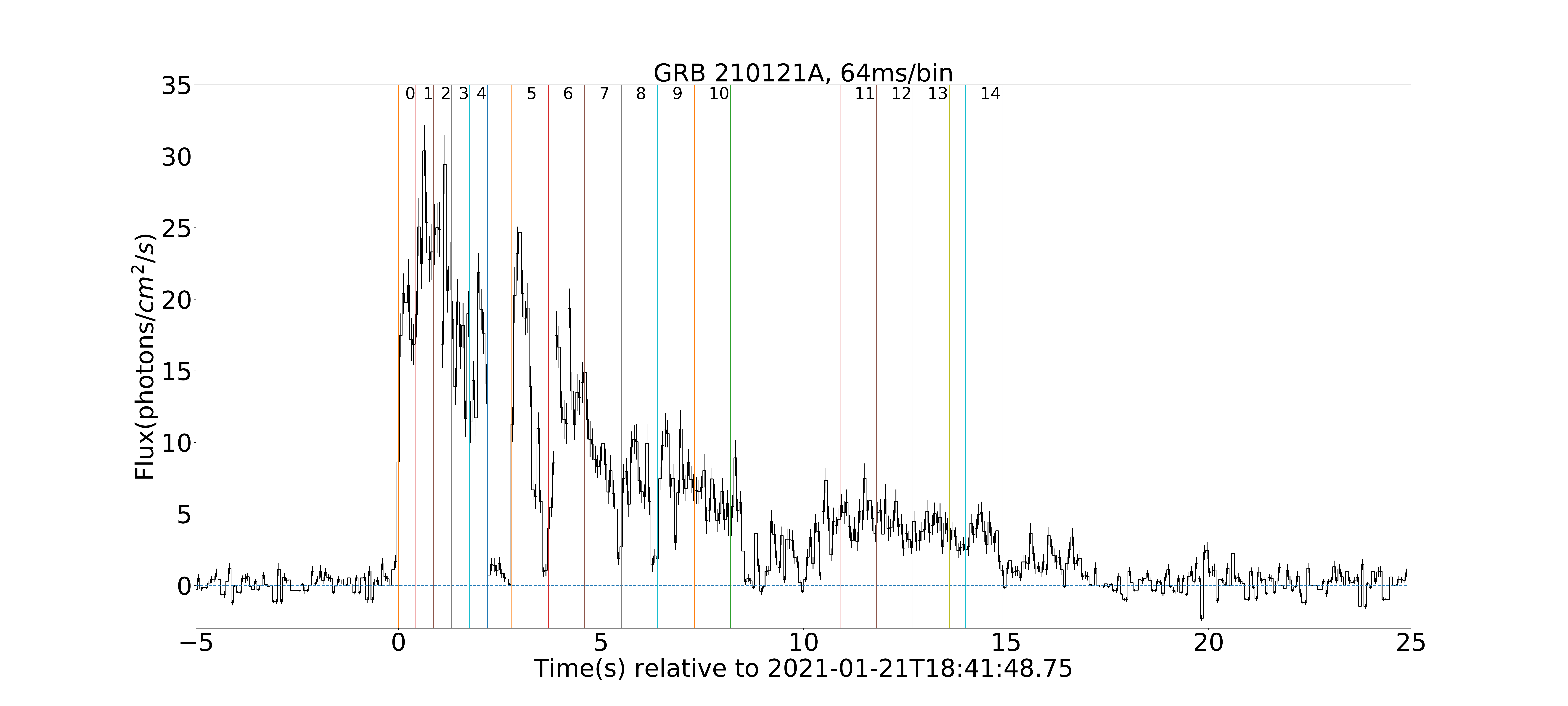}\\
\caption{\label{fig:phflux_GRB210121A}The photons flux of GRB 210121A, and the bins of 0-14 are labeled.}
\end{center}
\end{figure*}

GRB 210121A is observed by Insight-HXMT~\citep[GCN:][]{2021GCNHXMT}, Gravitational wave high-energy Electromagnetic Counterpart All-sky Monitor (GECAM)~\citep[GCN:][]{2021GCNGECAM}, Fermi Gamma-ray Burst Monitor (Fermi/GBM) on Jan 21st 2021. It triggered Insight-HXMT at 2021-01-21T18:41:48.750 UTC, and GECAM at 2021-01-21T18:41:48.800 UTC. The former one is taken to be the $T_{0}$ in the following analysis. The photon flux of GRB 210121A is shown in Figure~\ref{fig:phflux_GRB210121A}, which is extracted from the joint analysis utilized in \cite{song2022insighthxmt}. 
The cutoff power law (CPL) model is preferred for almost all the time-resolved slices as shown in \cite{Wang_2021}. The values of $\alpha$ of the first epoch obtained from time-resolved analysis are greater than -2/3, as well as those of the most part of the second epoch from $T_{0}$+2.8s to $T_{0}$+14.9s. According to \cite{10.1093/mnras/stab3132}, the NDP spectra from a hybrid outflow with a moderate magnetization have a larger high-energy photon index ($\beta$), and are more compatible with the BAND function rather than the CPL model. Therefore, we prefers a pure hot fireball (or the Poynting flux completely thermalized below the photosphere) in the prompt phase. 

\textbf{We can not justify the radiation mechanism only from $\alpha$ ~\citep{2020NatAs...4..174B,2018ApJ...860...72M}, thus a physical model fitting is needed.}

\section{The modeling of photospheric emissions from different regimes}\label{sec:NDPmodel}
 \textbf{A structured jet is supported by observation, \cite[e.g., studies on GRB 170817A,][]{2017MNRAS.471.1652L,2018MNRAS.475.2971B,2019ApJ...877L..40G} and the relativistic magnetohydrodynamic simulations for the GRB jet are also performed \cite[e.g., ][]{2019MNRAS.484L..98K}. In this analysis, we assume a structured jet with an opening angle $\theta_{\rm c}$ and luminosity $L_0$. Motivated by the results of \cite{1999ApJ...524..262M} and \cite{2003ApJ...586..356Z} e.g., 
  the jet is structured with an inner-constant and outer-decreasing angular baryon loading parameter profile  with the form for the GRB prompt emission phase~\citep{2001Gamma,10.1046/j.1365-8711.2002.05363.x, 2002Gamma,2003THE}
\begin{equation}\label{eta_theta}
(\eta(\theta) -\eta _{\min })^{2}=\frac{(\eta_{0}-\eta _{\min })^{2}}{%
(\theta /\theta _{c})^{2p}+1},
\end{equation}%
where $\eta$ is the angle-dependent baryon loading parameter which is also the bulk Lorentz factor $\Gamma$ in the saturated acceleration regime; $\eta_0$ is the maximum $\eta$, and also denoted as $\Gamma_0$; $\theta$ is the angle measured from the jet axis; $\theta_{c}$ is the half-opening angle for the jet core; $p$ is the power-law index of the profile; $\eta _{\min }=1.2$ is the minimum value of $\eta$, differing from unity for numerical reasons. The exact value of $\eta _{\min}$ only affects the very
low energy spectrum, many orders of magnitude below the observed peak
energy~\citep{2013A}. This baryon loading parameter profile have used in \cite{2013A}, \cite{2018ApJ...860...72M} and  \cite{2019The}.}

Flux of observed energy $E_{\rm obs}$ at the observer time $t$ in the case of the continuous wind is deduced and explained in Section~\ref{sec:NDPI}. The angle-dependent photosphere radius $R_{\rm ph}$, as the radius from which the optical depth for a photon that propagates in the radial direction is equal to unity, is defined as
\begin{equation}\label{Rph}
R_{\text{ph}}=\\
\begin{split}
&\begin{cases}
\left(\frac{\sigma _{\text{T}}}{6m_{\text{p}}c}%
\frac{d\dot{M}}{d\Omega }r_{0}^{2}\right) ^{1/3}\text{, } R_{\rm ph} \ll R_{\rm s} \text{, }\\
\left(\frac{\sigma _{\text{T}}}{2m_{\text{p}}c}%
\frac{d\dot{M}}{d\Omega }r_{0}^{2}\right) ^{1/3}\text{, } R_{\rm ph} \lesssim R_{\rm s} \text{, } \\
\frac{1}{(1+\beta )\beta \eta ^{2}}\frac{\sigma _{\text{T}}
}{m_{\text{p}}c}\frac{d\dot{M}}{d\Omega } \text{, }   R_{\rm ph} \geq R_{\rm s},\\
\end{cases}
\end{split}
\end{equation}
where $\beta$ is the velocity, $r_0$ is the radius of the central engine and $d\dot{M}(\theta )/d\Omega =L_0/4\pi c^{2}\eta (\theta )$ is the angle-dependent mass outflow rate per solid angle; \textbf{$m_{\rm p}$ is the mass of the proton, $c$ is the light speed, and $\sigma_{\rm T}$ is electron Thomson cross section.} Note the unsaturated regime used in \cite{Wang_2021} is $R_{\rm ph}\ll R_{\rm s}$. In this paper, an unsaturated emissions of $R_{\rm ph}\lesssim R_{\rm s}$ is considered and added between the case of $R_{\rm ph} \ll R_{\rm s}$ and $R_{\rm ph} \geq  R_{\rm s}$~\citep{2022MNRAS.512.5693S}.

$R_{\rm s}=\eta(\theta)r_0$ monotonically decreases
with $\theta$, while $R_{\rm ph}$ monotonically increases with $\theta$. Thus, there exists a critical value $\theta_{\rm cri}$, and for $\theta\geq\theta_{\rm cri}$, it satisfies $R_{\rm s}(\theta)\leq R_{\rm ph}(\theta)$, namely,
\begin{equation}\label{eq:ip}
\eta(\theta)\leq(\frac{\sigma_{\text{T}}}{2 m_{\text{p}}c}\frac{L_0}{4\pi c^2 r_0})^{1/4}.
\end{equation}
Note that for the intermediate photosphere, $\theta_{\rm cri}$ satisfies $0<\theta_{\rm cri}<5/\Gamma_0$, and the contribution from the saturated regime contributes to the prompt emission~\citep{2013A}, or the unsaturated emission is dominant. For an intermediate photosphere, $R_{\rm ph}$ is described by two forms: $R_{\rm ph}$ in lower latitude of the unsaturated part is described by the second item of Equation~(\ref{Rph}), while that of higher latitude takes the saturated form. 
 If for a set of parameters, $\theta_{\rm cri}$ is found to be 0 by Equation~(\ref{eq:ip}), the saturated regime works for all over the jet profile. Therefore in our analysis, the regime could be determined by the parameters obtained from the fitting to spectra.


\section{methods for data analysis}\label{sec:TRana}
\subsection{Binning method of light curves for time-resolved spectra} \label{sec:Binning}

In this work, the Bayesian blocks (BBlocks) method introduced by \cite{2013ApJ...764..167S} and suggested by \cite{2014On}, is applied with a false alarm probability $p_{0}=0.01$ on light curves. In some cases, the blocks are coarse for fine time-resolved analysis. \cite{2014On} suggested that the constant cadence (CC) method is accurate when the cadence is not too coarse. Therefore, we take a combination of BBlocks and CC methods, fine binning of constant cadence are performed in each block, and only the bins with the signal-to-noise ratio (S$/$N)~$\geq$~20 at least in one detector should be utilized. 15 bins from [$T_{0}$-0.01, $T_{0}$+ 14.90] s are shown and labeled in Figure~\ref{fig:phflux_GRB210121A}, where 5 bins are in the first epoch and the others are in the second one.  

\subsection{Background estimation and spectral fitting method}
A polynomial is applied to fit all the energy channels and then interpolated into the signal interval to yield the background photon count estimate for GRB data. 
Markov Chain Monte Carlo (MCMC) fitting is performed to find the parameters with maximum Poisson likelihood. The Sampling tool is $emcee$~\citep{2013PASP..125..306F}. \cite{2016MNRAS.463.1144W} suggested a bayesian information criterion (BIC) as a tool for model selection, a model that has a lower BIC value than the other is preferred. If the change of BIC between these two models, $\Delta$BIC is from 2 to 6, the preference for the model with the lower BIC is positive; if $\Delta$BIC is from 6 to 10, the preference
for that is strong; and if $\Delta$BIC is above 10, the preference is very strong.

\section{fit results}\label{sec:NDPfit}
In this section, time-averaged spectral fitting for two epochs as well as time-resolved analyses for fine bins are performed with the NDP model. The properties of the jet and the evolution of the outflow are extracted.
\subsection{ time-averaged results of two epochs }
\begin{figure*}
\begin{center}
 \centering
  \includegraphics[width=0.5\textwidth]{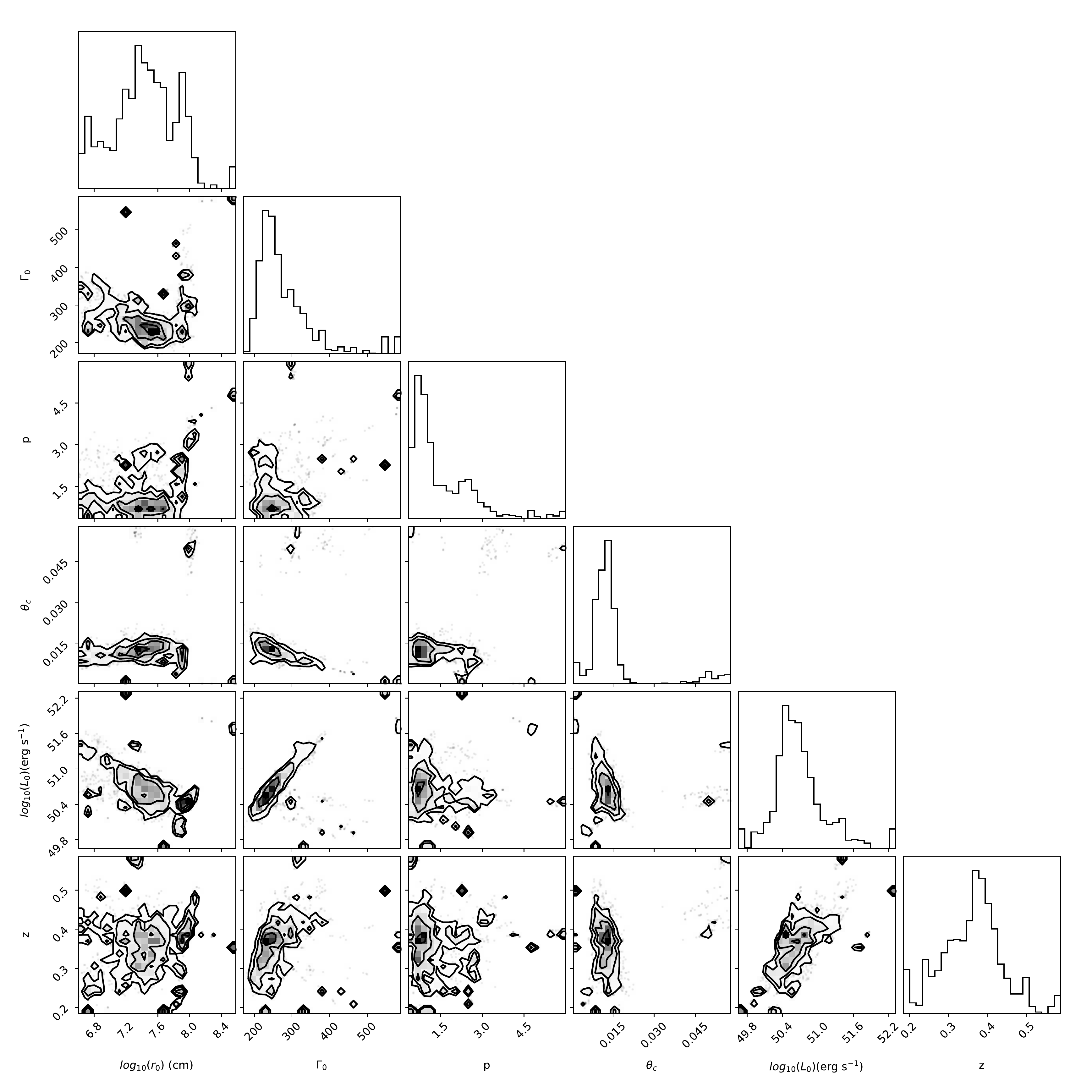}
  \put(-110,240){(a)epoch 1}
  \put(-230, 255){$log(r_0)$}
  \put(-185, 215){$\eta_0$}
  \put(-150, 175){$p$}
  \put(-115, 135){$\theta_{\rm c}$}
  \put(-80, 100){$log(L_0)$}
  \put(-25, 60){$z$}
   \includegraphics[width=0.5\textwidth]{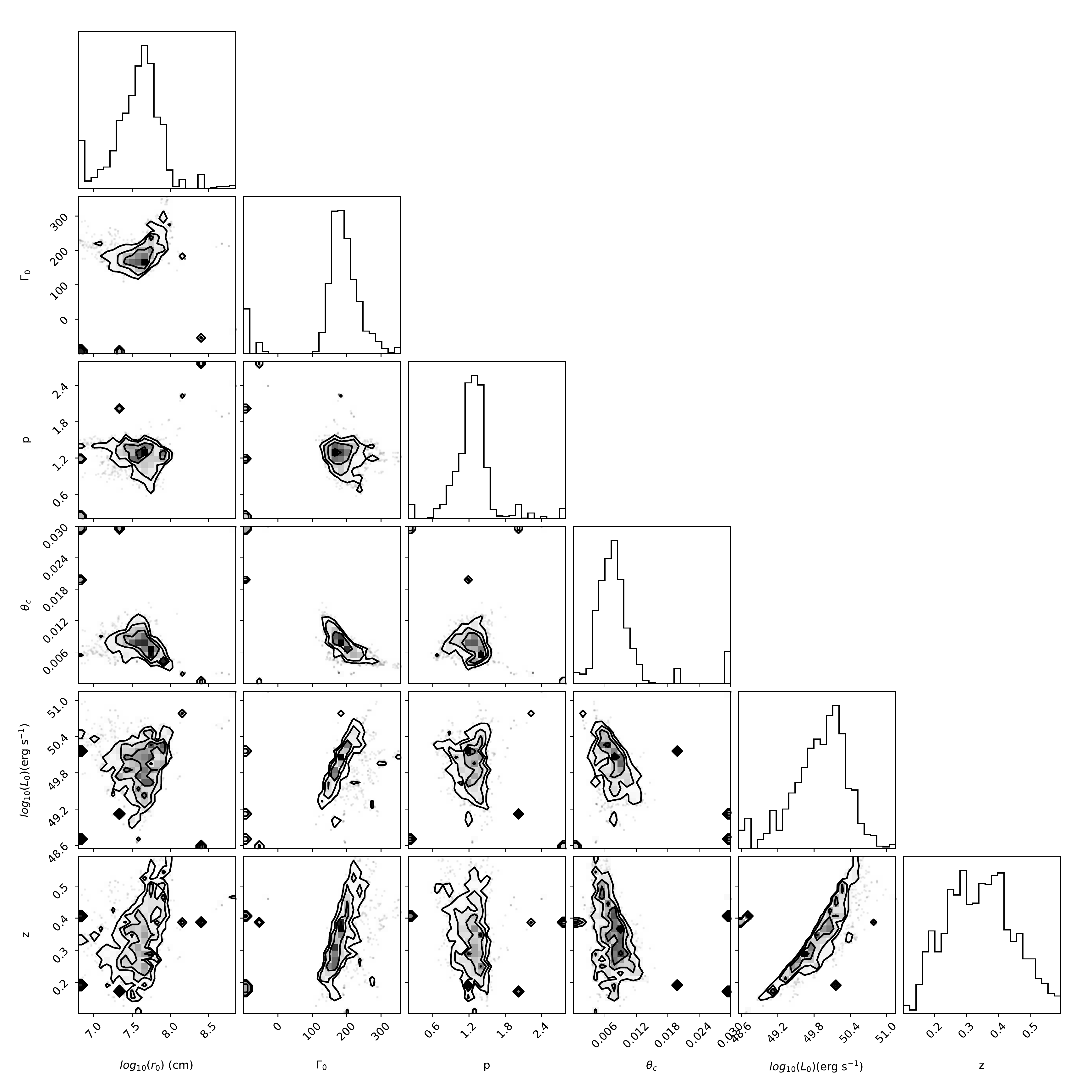}\put(-110,240){(b)epoch 2}
   \put(-230, 255){$log(r_0)$}
  \put(-185, 215){$\eta_0$}
  \put(-150, 175){$p$}
  \put(-115, 135){$\theta_{\rm c}$}
  \put(-80, 100){$log(L_0)$}
  \put(-25, 60){$z$}\\
   \includegraphics[width=0.5\textwidth]{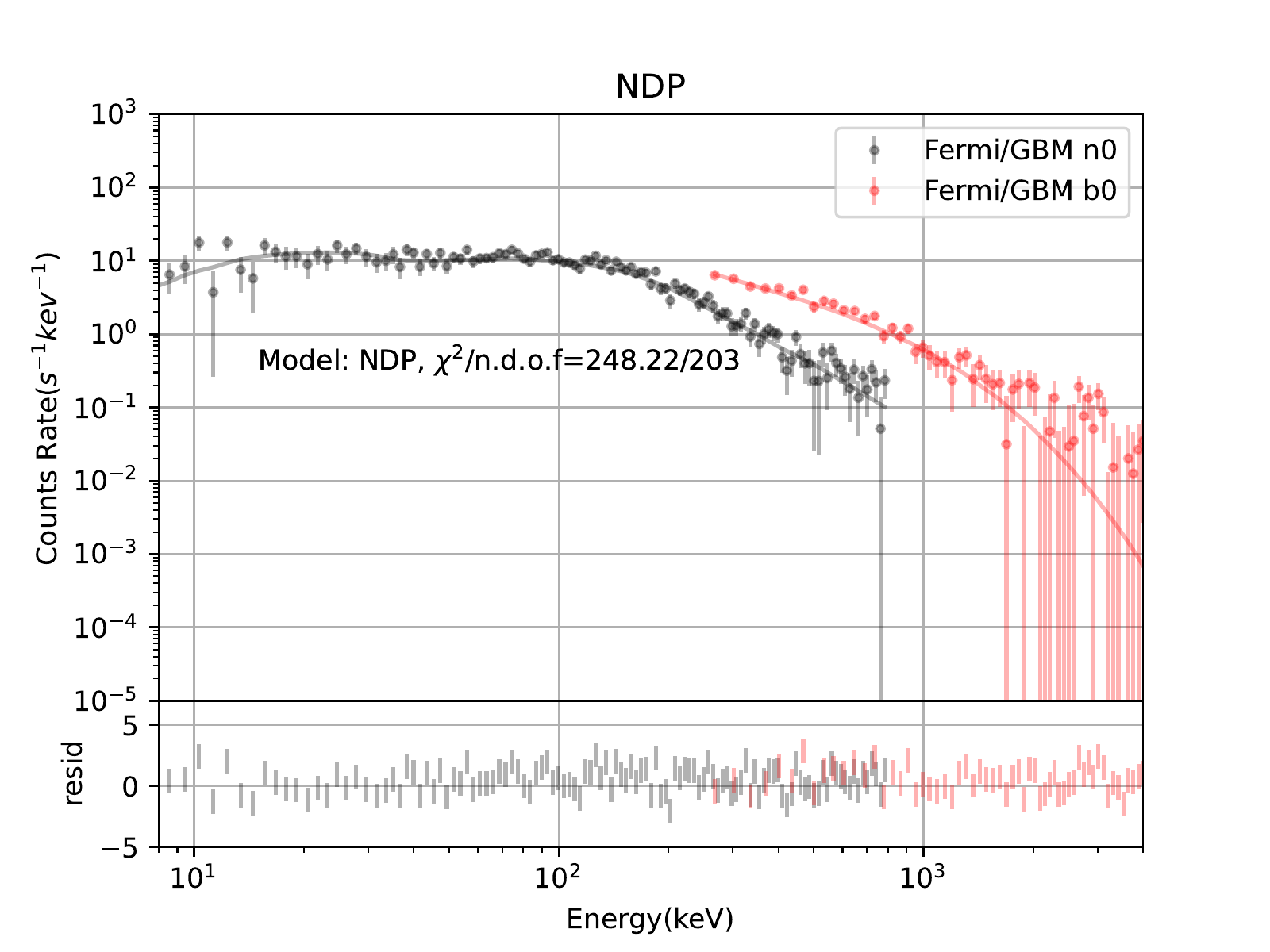}\put(-140,160){(c)epoch 1}
   \includegraphics[width=0.5\textwidth]{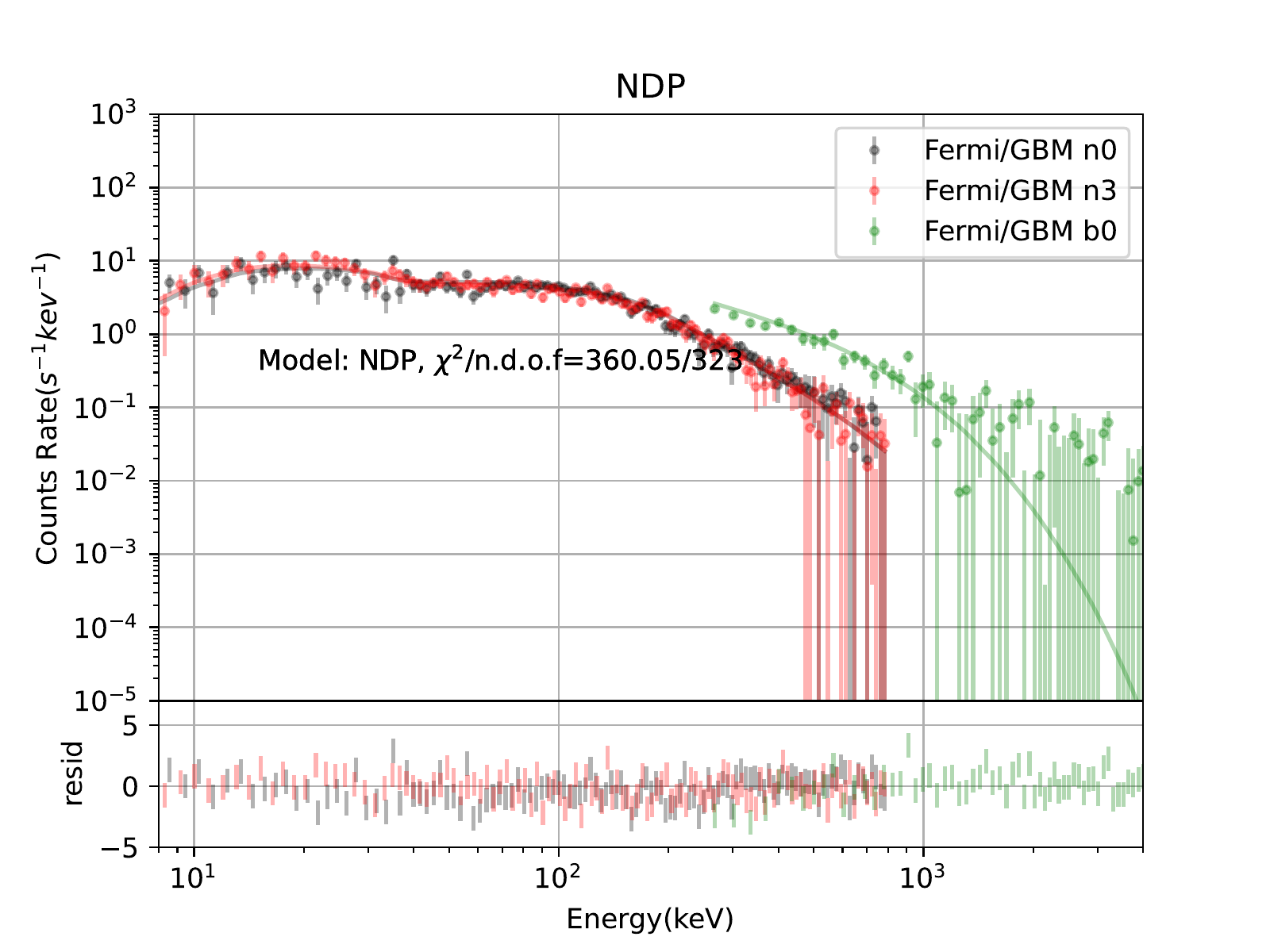}\put(-140,160){(d)epoch 2}   
\caption{(a) and (b) are the MCMC samples of fits results of two pulses of GRB 210121A. The plots for distributions of MCMC samples are generated by $corner$~\citep{corner}. (c) and (d) are the spectra and fit results for two pulses. \label{fig:MCMCsamples_GRB210121A} }
\end{center}
\end{figure*}

 Table~\ref{tab:fitresNDP_table_210121A} and  Figure~\ref{fig:MCMCsamples_GRB210121A} show the fit results, MCMC samples and spectra for the two epochs. It is found that $\theta_{\rm c}$ is at 0.01 with small uncertainties for both epochs. The values of $r_0$ are well consistent with $10^{7.61}$ cm. $p$ becomes larger in epoch 2 than that in epoch 1. The luminosity of outflow changes with time, and becomes smaller in epoch 2 than that in epoch 1, which almost has a similar trend to the flux. For the first epoch, $\chi^2=$248.2 and BIC=262.14 with the degree of freedom (dof) of 203, while $\chi^2$=360.0 and BIC=375.15 with the dof of 323 in the second epoch. We also performed the fit with an on-axis NDP model dominant by the regime of $R_{\rm ph}\ll R_{\rm s}$. However, the fit is not good, with BIC equals to 904.85 and 2014.26 for the two epochs with the same degrees of freedom. Thus, for on-axis observation, the NDP model in the regime of $R_{\rm ph}\ll R_{\rm s}$ is not preferred, and inconsistent with the spectra. 

These two results corresponds to $\theta_{\rm cri}=0$, which means the emission from the saturated regime are dominant. If considering with changes on $L_0$ and $\eta_0$ during the epochs, the emissions may be from the intermediate photosphere, or the unsaturated regime, which are determined in the fine time-resolved results.
\startlongtable
\setlength{\tabcolsep}{0.10em}
\begin{deluxetable*}{l|ccccccccc}
\tablewidth{0pt}
\tablecaption{\label{tab:fitresNDP_table_210121A} Fit results with on-axis NDP model of epoch 1 and 2. \textbf{In the time-resolved results from bin 0-14, the notations of `-' represent the fixed values in the fitting: $p=1.0$ for bins 0-4 in the first epoch and $p=1.27$ for bins 5-14 for the second epoch. For both epochs, log$r_0$=7.61, $\theta_{\rm c}=0.01$, z=0.37.}}
\tablehead{ \colhead{time bins(s)} &\colhead{log$(r_0($cm$))$} & \colhead{$\eta_0$}  &\colhead{p} &\colhead{$\theta_c$} &\colhead{log$(L_0$(erg s$^{-1}$))} &\colhead{z} &\colhead{BIC} &\colhead{$\frac{\chi^2}{dof}$}
&\colhead{$\theta_{\rm cri}(10^{-3})$}
}
\startdata
epoch1:[-0.01,2.19] &7.43$^{+0.43}_{-0.44}$ &268.1$^{+66.8}_{-34.6}$ &1.07$^{+1.44}_{-0.43}$ &0.010$^{+0.003}_{-0.003}$ &50.62$^{+0.42}_{-0.26}$ &0.37$^{+0.06}_{-0.09}$ &262.14  &$\frac{248.2}{203}$&0\\\hline
bin 0:[-0.01,0.43]& - &262.1$^{+10.2}_{-10.2}$ &- &- &50.42$^{+0.02}_{-0.02}$ &- &240.60 &$\frac{236.0}{196}$ &2.4\\
bin 1:[0.43,0.87]& - &341.8$^{+34.8}_{-23.2}$ &- &- &50.65$^{+0.02}_{-0.02}$ &- &258.76 &$\frac{254.2}{196}$&6.0\\
bin 2:[0.87,1.31]& - &280.2$^{+13.4}_{-4.7}$ &- &- &50.57$^{+0.03}_{-0.01}$ &- &314.46 &$\frac{309.9}{196}$ &1.7\\
bin 3:[1.31,1.75]& - &222.3$^{+3.4}_{-3.4}$ &- &- &50.53$^{+0.01}_{-0.02}$ &- &372.02 &$\frac{367.4}{196}$&0\\
bin 4:[1.75,2.19]& - &198.4$^{+6.0}_{-3.0}$ &- &- &50.46$^{+0.02}_{-0.02}$ &- &251.48 &$\frac{246.9}{196}$&0\\\hline\hline
epoch 2:[2.8-14.9] &7.61$^{+0.21}_{-0.31}$ &184.7$^{+44.17}_{-33.75}$ &1.27$^{+0.17}_{-0.25}$ &0.010$^{+0.003}_{-0.003}$ &50.15$^{+0.35}_{-0.50}$ &0.34$^{+0.11}_{-0.10}$ &375.15 &$\frac{360.0}{323}$&0\\\hline
bin 5:[2.80,3.70]& - &211.2$^{+3.1}_{-6.2}$ &- &- &50.35$^{+0.01}_{-0.01}$ &- &336.59 &$\frac{331.6}{316}$&0\\
bin 6:[3.70,4.60]& - &201.6$^{+3.0}_{-6.1}$ &- &- &50.37$^{+0.02}_{-0.02}$ &- &331.57 &$\frac{326.6}{316}$&0\\
bin 7:[4.60,5.50]& - &194.5$^{+8.1}_{-4.1}$ &- &- &50.06$^{+0.02}_{-0.02}$ &- &385.30 &$\frac{380.3}{316}$&0\\
bin 8:[5.50,6.40]& - &176.7$^{+9.6}_{-8.1}$ &- &- &50.01$^{+0.03}_{-0.02}$ &- &431.45 &$\frac{426.4}{316}$&0\\
bin 9:[6.40,7.30]& - &167.4$^{+2.9}_{-5.7}$ &- &- &50.04$^{+0.03}_{-0.03}$ &- &469.49 &$\frac{464.5}{316}$&0\\
bin 10:[7.30,8.20]& - &191.2$^{+9.5}_{-14.2}$ &- &- &49.97$^{+0.02}_{-0.03}$ &- &441.54 &$\frac{436.5}{316}$&0\\
bin 11:[10.90,11.80]& - &188.7$^{+14.2}_{-14.0}$ &- &- &49.80$^{+0.02}_{-0.02}$ &- &475.14 &$\frac{470.1}{316}$&3.0\\
bin 12:[11.80,12.70]& - &141.1$^{+2.5}_{-4.9}$ &- &- &49.81$^{+0.03}_{-0.05}$ &- &374.59 &$\frac{369.6}{316}$&0\\
bin 13:[12.70,13.60]& - &153.3$^{+5.9}_{-10.2}$ &- &- &49.78$^{+0.03}_{-0.03}$ &- &366.11 &$\frac{361.1}{316}$&0\\
bin 14:[14.00,14.90]& - &147.7$^{+7.2}_{-10.7}$ &- &- &49.69$^{+0.05}_{-0.03}$ &- &379.15 &$\frac{374.1}{316}$&0\\
\enddata
\end{deluxetable*}

\subsection{time-resolved results for fine bins}
\begin{figure*}
\begin{center}
 \centering
  \includegraphics[width=0.5\textwidth]{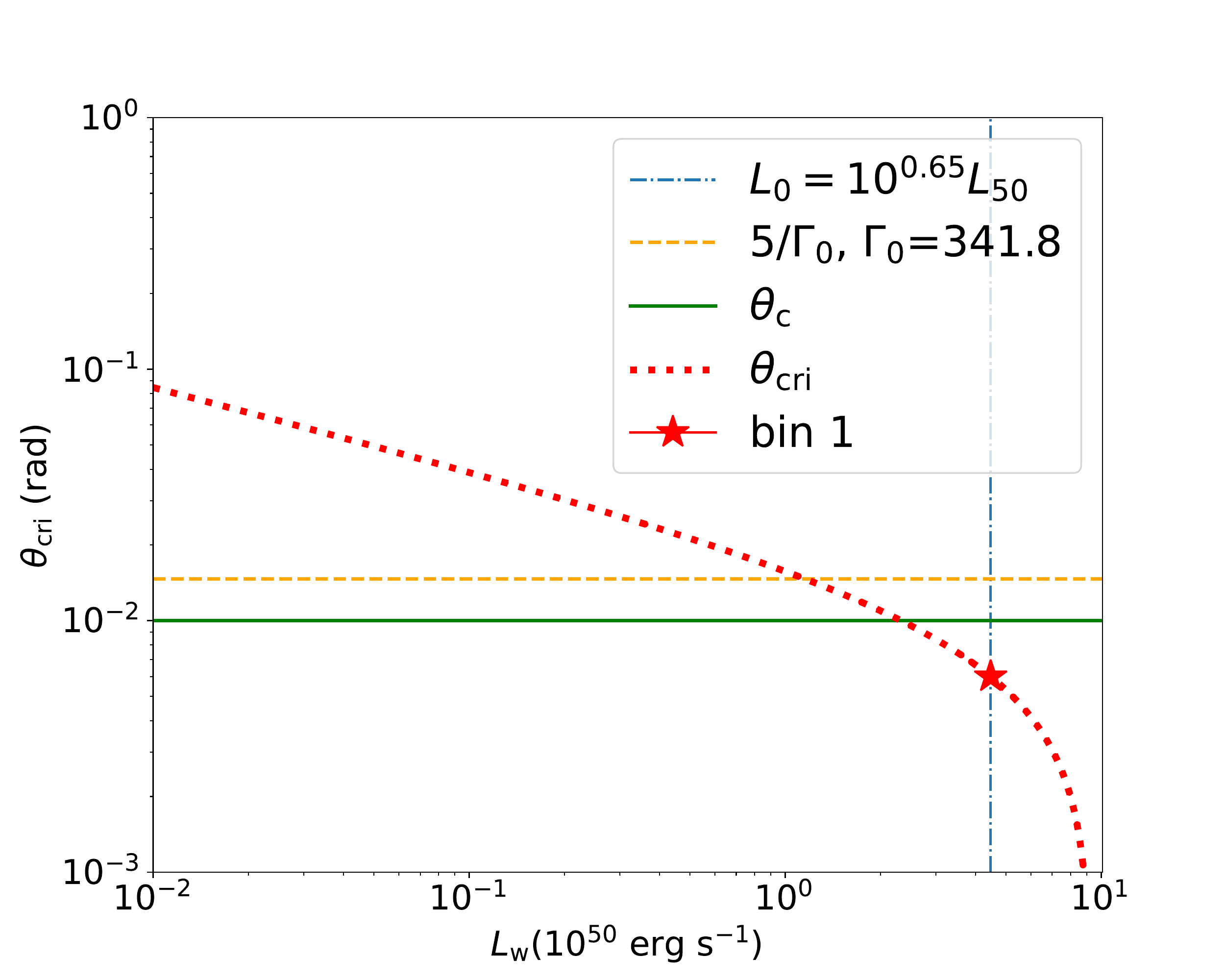}\put(-210,140){(a) bin 1}
  \includegraphics[width=0.5\textwidth]{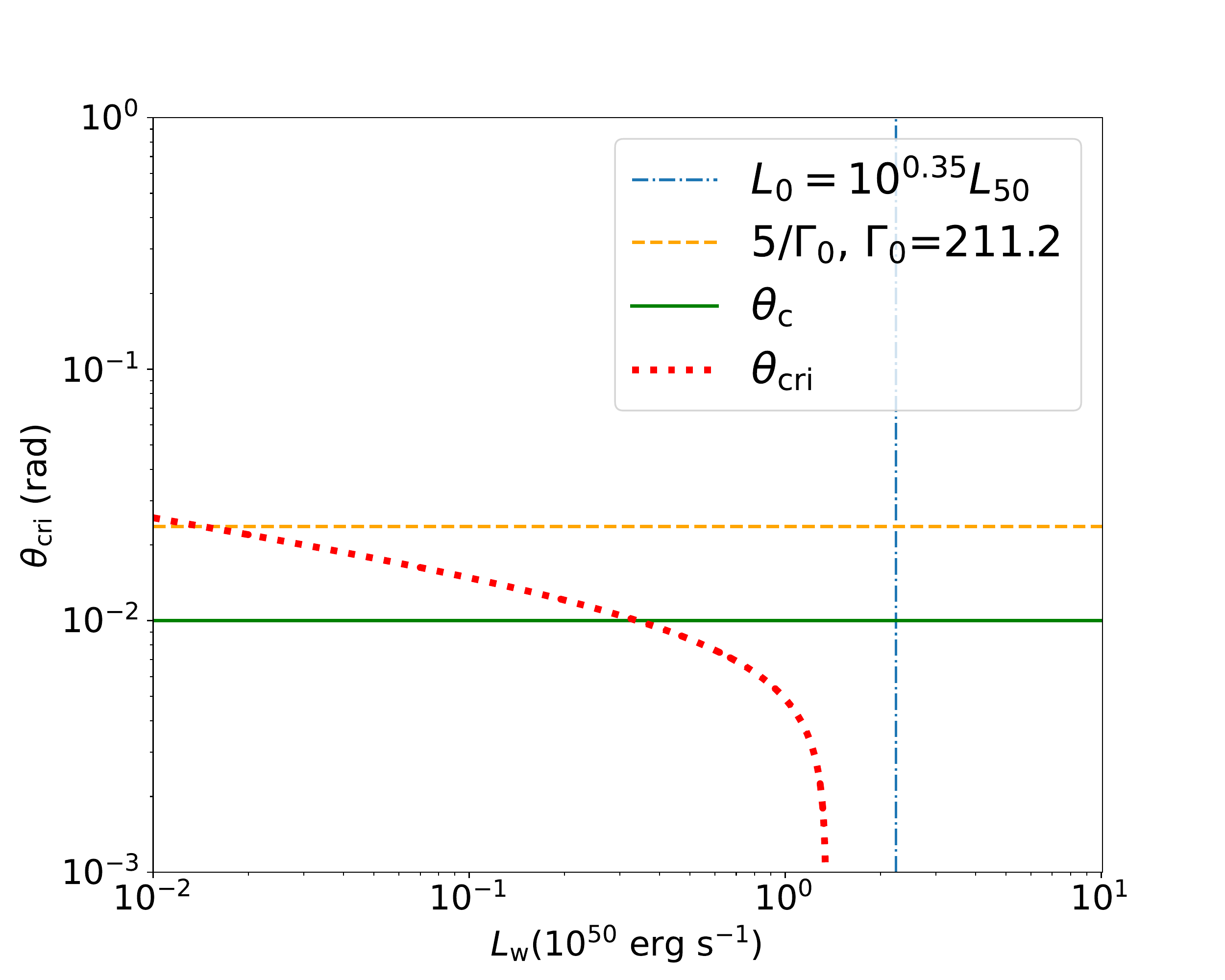}\put(-210,140){(b) bin 5}\\
\caption{The relations between $\theta_{\rm cri}$ and $L_0$ for bins 1 and 5. The red dotted lines represent the $\theta_{\rm cri}$ versus $L_0$, while the red stars represent the cases of the intermediate photonsphere. The orange dotted lines and green solid lines denote the values of $5/\Gamma_0$ and $\theta_{\rm c}$. The blue dot-dashed vertical lines represent the values of $L_0$ per bin.  \label{fig:regimeTR} }
\end{center}
\end{figure*}

\begin{figure*}
\begin{center}
 \centering
 \includegraphics[width=0.5\textwidth]{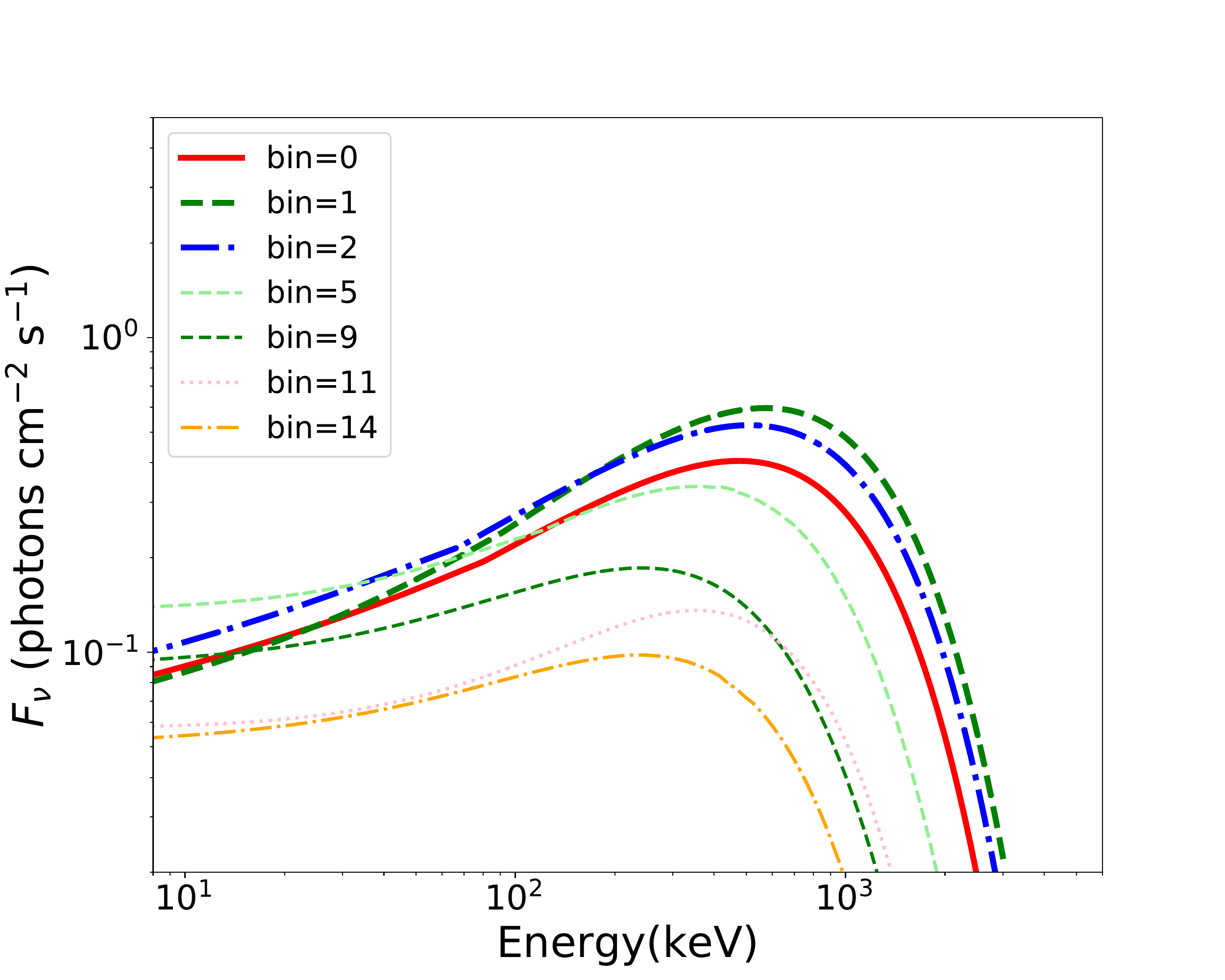}\put(-50,140){(a)}
  \includegraphics[width=0.5\textwidth]{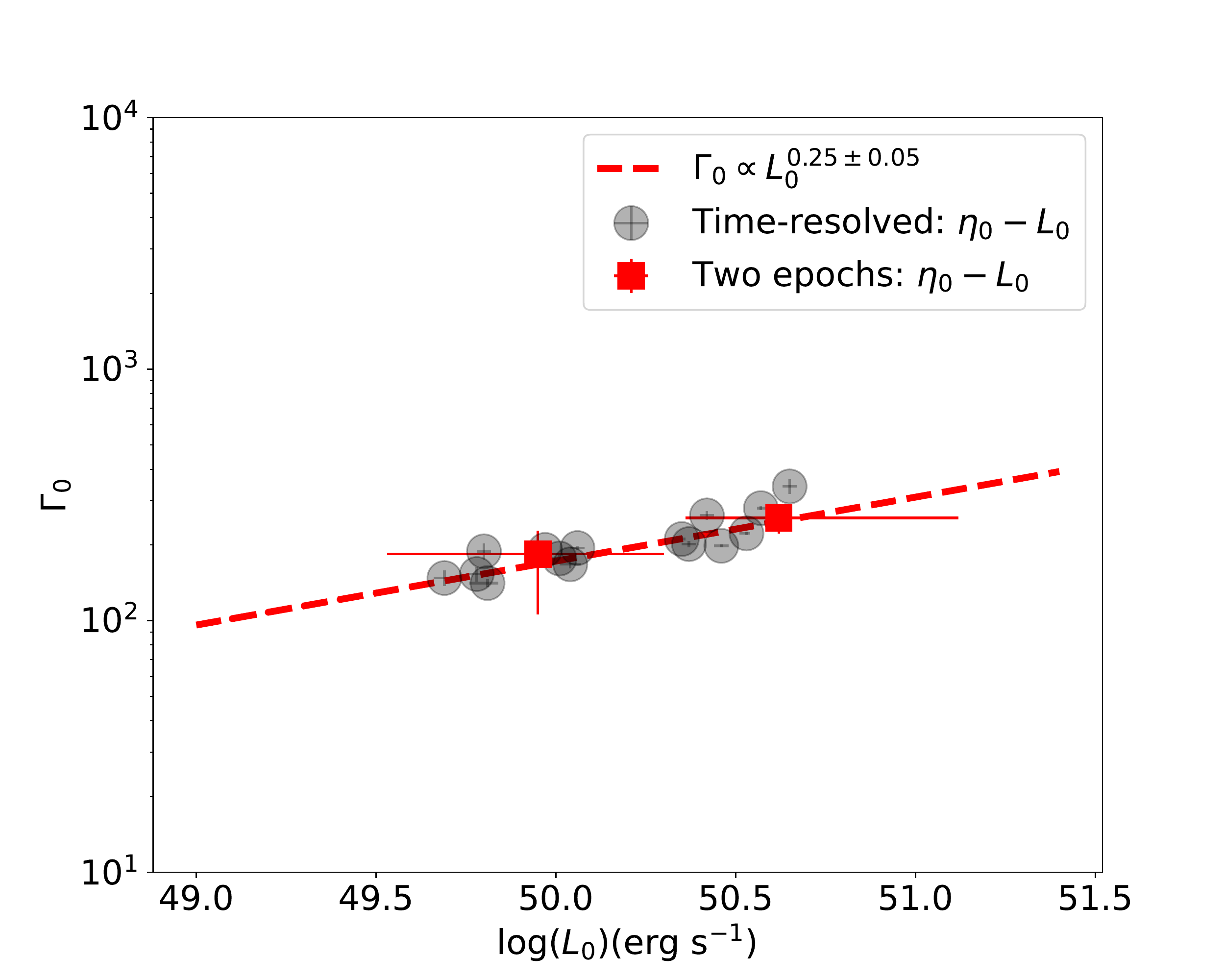}\put(-210,140){(b)}
\caption{(a) The time-resolved spectra of GRB 210121A of the two epochs.  (b) The correlation between $\Gamma_0$ and $L_0$. The black dots are from time-resolved fit results, while the red squares denotes the time-averaged fit results from the two epochs. The red dashed line denotes the fits result of $\Gamma_0$ and $L_0$.  \label{fig:etalumi_rela} }
\end{center}
\end{figure*}

 \textbf{Since the uncertainties of fit results from time-averaged spectra seem not large, we could assume that $r_0$, $p$, $\theta_{\rm c}$ have small changes during each epoch; $z$ stays constant for the whole GRB duration. Therefore, they could be fixed in the time-resolved fitting to suppress the uncertainty in extracting the correlation of $\Gamma_0-L_0$.} The time-resolved analysis is performed with float $\eta_0$ and $L_0$, while other parameters are fixed to $p=1.0$ for bins 0-4 in the first epoch and $p=1.27$ for bins 5-14 for the second epoch. For both epochs, log$r_0$=7.61, $\theta_{\rm c}=0.01$, z=0.37. The fit results are listed in Table~\ref{tab:fitresNDP_table_210121A}, where four bins have non-zero $\theta_{\rm cri}$.  All over the bins, bin 1 has the largest $\eta_0$ and flux, as well as the largest $\theta_{\rm cri}$. We can conclude that it has larger contribution from the unsaturated regime than the other bins. The relation of $\theta_{\rm cri}$ and $L_0$ for each set of ($r_0$, $p$, $\theta_{\rm c}$) of bin 1 and bin 5 are shown in Figure~\ref{fig:regimeTR} in red dotted lines. The values of luminosity measured in time-resolved analysis are represented by vertical dot-dashed lines in blue. As shown in Figure~\ref{fig:regimeTR}~(a), the corresponding $\theta_{\rm cri}$ of bin 1 denoted as a red star is less than $\theta_{\rm c}$ (denoted by the green solid line) as well as $5/\Gamma_0$ (denoted by the orange dashed line), which means that the emission is from the intermediate photosphere. For comparison, 
 emissions in other bins, e.g., bin 5, is shown in Figure~\ref{fig:regimeTR}~(b). The blue vertical line does not intersect with the red dotted line representing $\theta_{\rm cri}$, which means the corresponding $\theta_{\rm cri}$ is 0, and the emission is from the saturated regime. The time-resolved spectra are shown in Figure~\ref{fig:etalumi_rela}~(a), where the luminosity and hardness decrease by time generally.
 
 The relations between $\eta_0$ and $L_0$ are shown in Figure~\ref{fig:etalumi_rela}. $\Gamma_0\propto L^{0.25\pm0.05}_0$ is extracted from the time-resolved results denoted by black dots. The time-averaged results for two epochs represented by red squares are also plotted, and consistent well with this trend. 

\section{Discussion and conclusion}\label{sec:discuss}
In the previous study, \cite{Wang_2021} performed a fit on the first epoch in GRB 210121A with the NDP model in unsaturated regime ($R_{\rm ph} \ll R_{\rm s}$), and the observation is determined to be greatly off-axis with the extracted $\theta_{\rm v}\simeq 15/\Gamma_0$. In this paper, we use NDP model with considering the intermediate photospheric emissions to describe two epochs of GRB 210121A and give an alternative description with an on-axis observation. The obtained luminosity, log$(L_0($erg s$^{-1}))=50.62^{+0.42}_{-0.26}$, is naturally smaller than that of off-axis, log$(L_0($erg s$^{-1}))={51.94}^{+0.84}_{-0.50}$ (this value is from~\cite{Wang_2021}), and the uncertainties of this result is smaller as well. The extracted redshift in this work, $z\sim0.37^{+0.06}_{-0.09}$ is also consistent with the prediction of photosphere death line, $z\sim$[0.3, 3.0]~\citep{Zhang_2012, Wang_2021}.

  \textbf{The intermediate photosphere always has a moderate $\alpha$ between these two regimes~\citep{2022MNRAS.512.5693S}, where $\alpha$ is extracted from spectra below the peak energy with an exponential cut-off power law.} This could explain that bin 1 has the largest $\alpha=-0.22^{+0.07}_{-0.07}$ (this value is from~\cite{Wang_2021}), while the others has smaller $\theta_{\rm cri}$ and more contribution from the saturated regimes. According to \cite{Wang_2021}, it seems that the off-axis unsaturated NDP model could also give a smaller $\alpha$ than on-axis unsaturated NDP model, therefore, there may be double solutions for the observation, and the on-axis NDP model with considering intermediate photonsphere gives an alternative solution.

  \cite{2010ApJ...725.2209L} presented a correlation of $\Gamma_0\propto E^{0.25}_{\rm iso, \gamma}$. 
 \cite{L_2012} discovered an even tighter correlation $\Gamma_0\propto L^{0.30}_{\rm iso, \gamma}$ from 50 GRBs and proposed an interpretation. Considering the beaming factor ($f_{\rm b}\propto L^{-0.145}_{\rm iso, \gamma}$) and the similar efficiency of jet kinetic energy via internal shock in dissipation to prompt emissions of GRBs, it has $\Gamma_0\propto L^{0.22}_{\rm iso, \gamma}$. However, \cite{YI20171} found that the data are more consistent with the latter mechanism, the index$\sim 0.14$.  In this analysis, $\Gamma_0-L_0$ of GRB210121A is obtained from the time-resolved analysis, and there are some advantages. First, we do not need to consider the efficiency of the prompt emission, and  $L_0\propto\dot{E}$ ($\dot{E}$ is the power). Second, we obtain the baryon loading parameter directly from the the fit result, and consider the regime in this procedure. The Lorentz factor of the outflow may be lower than the baryon loading parameter if the emission is from the unsaturated regime. $\eta_0\propto L_0^{0.25\pm0.05}$ is extracted from time-resolved analysis, and it is more consistent with the mechanism of $\nu\overline{\nu}$ annihilation. 
 
 In summary, the NDP model with considering the intermediate photosphere is first extracted in \textbf{GRB 210121A}, and it could well explain the changes on $\alpha$ during the pulses. The correlation of $\eta_0-\Gamma_0$ is extracted from the time-resolved analysis, implies the jet of GRB 210121A may be mainly from the $\nu\overline{\nu}$ annihilation.
\acknowledgements {
 The authors thank supports from the National Program on Key Research and Development Project (2021YFA0718500). The authors are very grateful to the GRB data of Fermi/GBM, HXMT and GECAM. We are very grateful for the comments and suggestions of the anonymous referees. Xin-Ying Song thanks Dr Yan-Zhi Meng for his suggestion on the algorithm of spectra fitting procedure and  Prof. Wen-Xi Peng for his support during the work.
}



\appendix
\section{APPENDIX}
\subsection{The flux of photospheric emissions from different regimes }\label{sec:NDPI}

As discussed in \cite[e.g.]{2013A, Deng_2014,2018ApJ...860...72M}, flux of observed energy $E_{\rm obs}$ at the observer time $t$ in the case of continuous wind could be shown as in Equation~(\ref{FEob}),
\begin{equation} \label{FEob}
\begin{split}
 F_{E}^{\rm obs}(\theta _{\rm{v}}, E_{\rm obs}, t) &=\frac{1}{4\pi d_{\text{L}}^{2}}
\int\int (1+\beta )D^{2}\frac{d\dot{N}_{\gamma }}{d\Omega }\times \frac{R_{
\text{ph}}}{r^{2}}\\
&\exp \left( -\frac{R_{\text{ph}}}{r}\right)  \times \left\{ E\frac{dP}{dE}\right\}\times \frac{\beta c}{u} d\Omega dr, \\
&r=\frac{\beta c t}{u}, E=E_{\rm obs}(1+z),
\end{split}
\end{equation}
where the velocity $\beta=\frac{v}{c}$ and the Doppler factor $D=[\Gamma (1-\beta\cos \theta _{\text{LOS}})]^{-1}$ both depend on the angle $\theta$ to the jet axis of symmetry, in which $\theta _{\text{LOS}}$ is the angle to the line of sight (LOS) of the observer. The viewing
angle $\theta _{\text{v}}$ is the angle of the jet axis of symmetry to the LOS. \textbf{If it is on-axis, we have $\theta _{\text{v}}=0$ and $\theta _{\text{LOS}}=\theta$.} $d\dot{N}_{\gamma }/d\Omega =\dot{N}_{\gamma }/4\pi $ and $\dot{N}%
_{\gamma }=L/2.7k_{\text{B}}T_{0}$, where $L$ is the total outflow luminosity, $T_{0}=(L/4\pi r_{0}^{2}ac)^{1/4}$ is the base outflow temperature and $a$ is the radiation constant. 
 $dP/dE$ describes the probability for a photon to have an observer frame energy between $E$ and $E+dE$ within volume element $dV$, and it is a comoving Planck distribution
with the comoving temperature $T^{\prime}(r,\Omega)$, as shown in Equation~(\ref{dPdE}),
\begin{equation}\label{dPdE}
\frac{dP}{dE}=\frac{1}{2.40(k_{\text{B}}T_{\text{ob}})^{3}}\frac{E^{2}}{\exp
(E/k_{\text{B}}T_{\text{ob}})-1},
\end{equation}%
where $k_{\rm B}$ is the Boltzmann constant, $T_{\text{ob}}(r,\Omega )=D(\Omega )\cdot $ $T^{\prime }(r,\Omega )$
is the observer frame temperature, and in the saturated acceleration regime ( $R_{\rm s}<R_{\rm ph}$), it is defined by~\citep{2000ApJ...530..292M,Deng_2014}
\begin{equation}
T^{\prime }(r,\Omega )=\left\{
\begin{array}{c}
\frac{T_{0}}{\Gamma (\Omega )},\text{ \ \ \ \ \ \ \ \ \ \ \ \ \ \ \ \ \ \ \ }%
r<R_{s}(\Omega )<R_{\text{ph}}(\Omega ), \\
\frac{T_{0}[r/R_{s}(\Omega )]^{-2/3}}{\Gamma (\Omega )},\text{ \ \ \ \ \ \ \
}R_{s}(\Omega )<r<R_{\text{ph}}(\Omega ), \\
\frac{T_{0}[R_{\text{ph}}(\Omega )/R_{s}(\Omega )]^{-2/3}}{\Gamma (\Omega )},%
\text{ }R_{s}(\Omega )<R_{\text{ph}}(\Omega )<r.%
\end{array}%
\right.
\end{equation}

In the unsaturated acceleration case,  comoving temperature $T^{\prime}(r)$ is given by
\begin{equation}\label{T2}
T^{\prime }(r)=T_{0}/\Gamma,
\end{equation}
where $\Gamma$ is $R_{\rm ph}/r_0$. The spectra are always obtained within a time interval in time-averaged or time-resolved analysis. The luminosity is taken to be constant as an averaged value, and the spectra is taken as a time-integrated spectra in this time interval. 

Continuous wind could be assumed to consist of many thin layers from an impulsive injection at its injection time $\hat{t}$, and the wind luminosity at $\hat{t}$ is denoted as $L_{\rm w}(\hat{t})$. The flux of observed energy $E_{\rm obs}$ at the observer time $t$ is given by 
\begin{equation} \label{eq:FEob2}
\begin{split}
 &F_{E_{\rm{obs}}}^{\rm obs}(\theta _{\rm{v}}, E_{\rm obs}, t)=\\
 &\begin{cases}
\int\nolimits_{0}^{t}F_{E}^{\text{obs}}(\theta _{\text{v}},t,\hat{t}, L_{\rm w}(\hat{t}))d%
\hat{t},\textbf{\ \ \  }t< t_{\rm D}, \\
\int\nolimits_{0}^{t_{\rm D}}F_{E}^{\text{obs}}(\theta _{\text{v}},t,\hat{t}, L_{\rm w}(\hat{t}))d%
\hat{t},\textbf{\ \ \  }t\geq t_{\rm D},\\
\end{cases}
\end{split}
\end{equation}
where $t_{\rm D}$ is the duration of emission of the central engine, and $r=\frac{\beta c (t-\hat{t})}{u}$ in $F_{E}^{\text{obs}}(\theta _{\text{v}},t,\hat{t}, L_{\rm w}(\hat{t}))$ in Equation~(\ref{FEob}). The observed GRB has a duration, thus, it is reasonable that the central engine produces a continuous wind. For the case of the constant wind luminosity, after the central engine has an abrupt shutdown ($t>t_{\rm D}$), the flux sharply drops. Moreover, the observed light curves of the GRBs show relatively slow change in luminosity, unlike the steep rise and fall (almost within $10^{-3}$ s), thus, a continuous wind with variable luminosity is used to simulate a GRB pulse.

\label{sec:pubcharge}


\bibliography{GRB210121Abib}{}
\bibliographystyle{aasjournal}



\end{document}